\documentclass{article}

\usepackage{fullpage}
\usepackage{amssymb,amsfonts,amsmath}
\usepackage{graphicx}
\usepackage{rotate}
\usepackage{epsfig}
\usepackage{cite}
\usepackage{color}
\usepackage{url}
\usepackage{multirow}
\usepackage{setspace}

\title{Virality Prediction and Community Structure in Social Networks}

\author{
Lilian Weng, Filippo Menczer, and Yong-Yeol Ahn~\footnote{Email of the corresponding author: \texttt{yyahn@indiana.edu}}\\
\\
Center for Complex Networks and Systems Research,\\
School of Informatics and Computing,\\
Indiana University, Bloomington, IN 47408, USA
}

\date{}

\begin{document}
\maketitle

\begin{abstract}

How does network structure affect diffusion? Recent studies suggest that the answer depends on the type of contagion. Complex contagions, unlike infectious diseases (simple contagions), are affected by social reinforcement and homophily. Hence, the spread within highly clustered communities is enhanced, while diffusion across communities is hampered. 
A common hypothesis is that memes and behaviors are complex contagions. We show that, while most memes indeed spread like complex contagions, a few viral memes spread across many communities, like diseases. 
We demonstrate that the future popularity of a meme can be predicted by quantifying its early spreading pattern in terms of community concentration. The more communities a meme permeates, the more viral it is.  
We present a practical method to translate data about community structure into predictive knowledge about what information will spread widely. This connection contributes to our understanding in computational social science, social media analytics, and marketing applications.

\end{abstract}
 
\newpage


Diseases, ideas, innovations, and behaviors spread through social 
networks\cite{newman_book, kleinberg_book, goffman1964nature, 
Daley1964rumour, Anderson1992book, Rogers:2003vn, vesp2008book, 
vespignani2001SISonSF, Centola2010SpreadExp, ugander2012diversity, 
bond-facebook-2012, christ2007obesity}.
With the availability of large-scale, digitized data on social communication\cite{Vespignani2009, Lazer2009}, the study of diffusion of \emph{memes} (units of transmissible information) has become feasible recently\cite{moreno2004dynamics,Leskovec2009Meme,borge2011structural,lehmann2012dynamical}.
The questions of how memes spread and which will go viral have recently attracted much attention
across disciplines, including marketing\cite{Rogers:2003vn, aral2011viral},
network science\cite{Leskovec2007viral, Weng:2012scirep}, 
communication\cite{Berger:2009viral}, and social media analytics\cite{Jamali2009digging,Szabo2010pred,Suh2010}. 
Network structure can greatly affect the spreading process\cite{Watts2002,Dodds2004,moreno2004dynamics}; 
for example, infections with small spreading rate persist 
in scale-free networks\cite{vespignani2001SISonSF}.  
Existing research has attempted to characterize viral memes in terms of message
content\cite{Berger:2009viral}, temporal variation\cite{Leskovec2009Meme,Szabo2010pred}, 
influential users\cite{Kitsak2010kcore,aral2011viral}, finite user attention\cite{Weng:2012scirep,
lehmann2012dynamical}, and local neighborhood structure\cite{ugander2012diversity}. 
Yet, what determines the success of a meme and how a meme interacts with the underlying network
structure is still elusive.  A simple, popular approach in studying meme
diffusion is to consider memes as diseases and apply epidemic models\cite{Daley1964rumour, goffman1964nature}.  
However, recent studies demonstrate that diseases and behaviors 
spread differently; they have therefore been referred to as \emph{simple} versus \emph{complex 
contagions,} respectively\cite{Centola2010SpreadExp, Backstrom2006GFL}.

Here we propose that network communities\cite{fortunato2010community,
Rosvall2008InfoMap, Ahn2010LinkCommunity}---strongly clustered groups of
people---provide a unique vantage point to the challenge of predicting viral
memes. We show that (i)~communities allow us to estimate how much the spreading
pattern of a meme deviates from that of infectious diseases; (ii)~viral memes
tend to spread like epidemics; and finally (iii)~we can \emph{predict} the
virality of memes based on early spreading patterns in terms of community
structure.
We employ the popularity of a meme as an indicator of its virality; viral memes 
appear in a large number of messages and are adopted by many people.

\begin{figure*}
\centering
\includegraphics[width=0.85\textwidth]{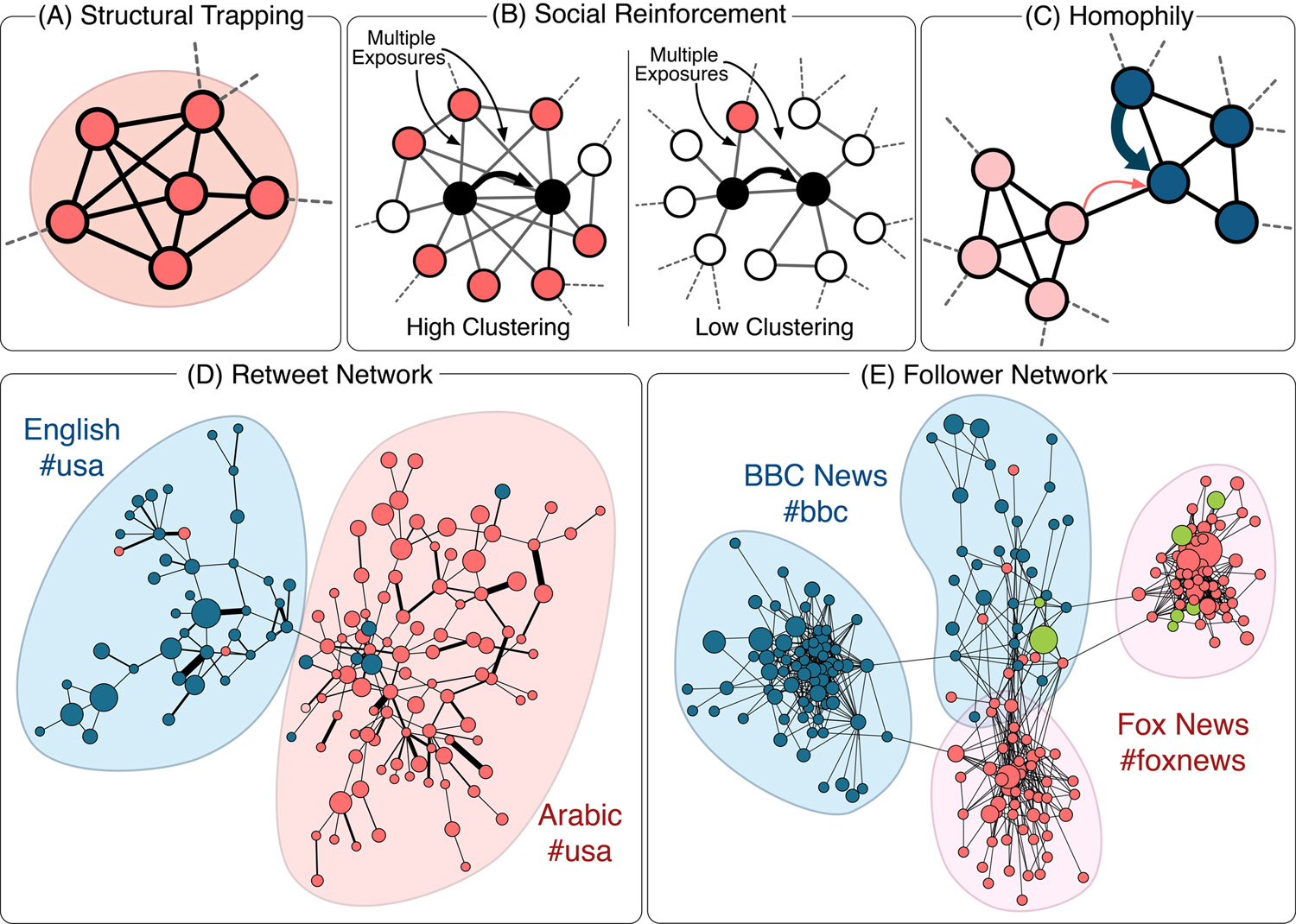}
\caption{ The importance of community structure in the spreading of social
    contagions.  (A)~\emph{Structural trapping}: dense communities with few
    outgoing links naturally trap information flow.  (B)~\emph{Social
	reinforcement}: people who have adopted a meme (black nodes) trigger
    multiple exposures to others (red nodes). In the presence of high
    clustering, any additional adoption is likely to produce more multiple
    exposures than in the case of low clustering, inducing cascades of
    additional adoptions.  (C)~\emph{Homophily}: people in the same community
    (same color nodes) are more likely to be similar and to adopt the same
    ideas.  (D)~Diffusion structure based on retweets among Twitter users sharing the hashtag
    \texttt{\#USA}. Blue nodes represent English users and red nodes are Arabic
    users. Node size and link weight are proportional to retweet
    activity.   (E)~Community structure among Twitter users sharing the
    hashtags \texttt{\#BBC} and \texttt{\#FoxNews}. Blue nodes represent
    \texttt{\#BBC} users, red nodes are \texttt{\#FoxNews} users, and users who
    have used both hashtags are green. Node size is proportional to usage
    (tweet) activity, links represent mutual following relations. 
    \label{fig:illust}
} 
\end{figure*}

Community structure has been shown to affect information diffusion, including global cascades\cite{galstyan2007cascading, gleeson2008cascades}, the speed of propagation\cite{JP2007pnas}, and the activity of individuals\cite{weak_ties_1973,grabowicz2012social}. One straight-forward effect is that communities are thought to be able to cripple the global spread because they act as traps for random flows\cite{weak_ties_1973,JP2007pnas} (Fig.~\ref{fig:illust}(A)). 
Yet, the causes and consequences of the \emph{trapping} effect have not been fully understood, 
particularly when \emph{structural} trapping is combined with two important phenomena: social
reinforcement and homophily. Complex contagions are sensitive to \emph{social
reinforcement}: each additional exposure significantly increases the chance of
adoption.  Although the notion is not new\cite{threshold_model}, it was only
recently confirmed in a controlled experiment\cite{Centola2010SpreadExp}. 
A few concentrated adoptions inside highly clustered
communities can induce many multiple exposures (Fig.~\ref{fig:illust}(B)).  The adoption of
memes within communities may also be affected by \emph{homophily}, according to
which social relationships are more likely to form between similar
people\cite{McPherson2001homophily, Centola2011homophily}.  Communities capture
homophily as people sharing similar characteristics naturally establish more
edges among them.  Thus we expect similar tastes among community members,
making people more susceptible to memes from peers in the same community
(Fig.~\ref{fig:illust}(C)).  Straightforward examples of homophilous communities are those
formed around language or culture (Fig.~\ref{fig:illust}(D,E)); people are much more likely to
propagate messages written in their mother tongue.  Separating social contagion
and homophily is difficult\cite{aral_distinguishing_2009,
Shalizi2012homophily}, and we interpret complex contagion broadly to include 
homophily; we focus on how both social reinforcement and homophily effects 
collectively boost the trapping of memes within dense communities, not on the 
distinctions between them.  

To examine and quantify the spreading patterns of memes, we analyze a dataset
collected from Twitter, a micro-blogging platform that allows millions of
people to broadcast short messages (`tweets').  People can `follow' others to
receive their messages, forward (`retweet' or ``RT'' in short) tweets
to their own followers, or mention (`@' in short) others in tweets. People
often label tweets with topical keywords (`hashtags').  We consider each
hashtag as a meme.

\section*{Results}

\subsection*{Communities and Communication Volume} 

\begin{figure}
\centering
\includegraphics[width=\columnwidth]{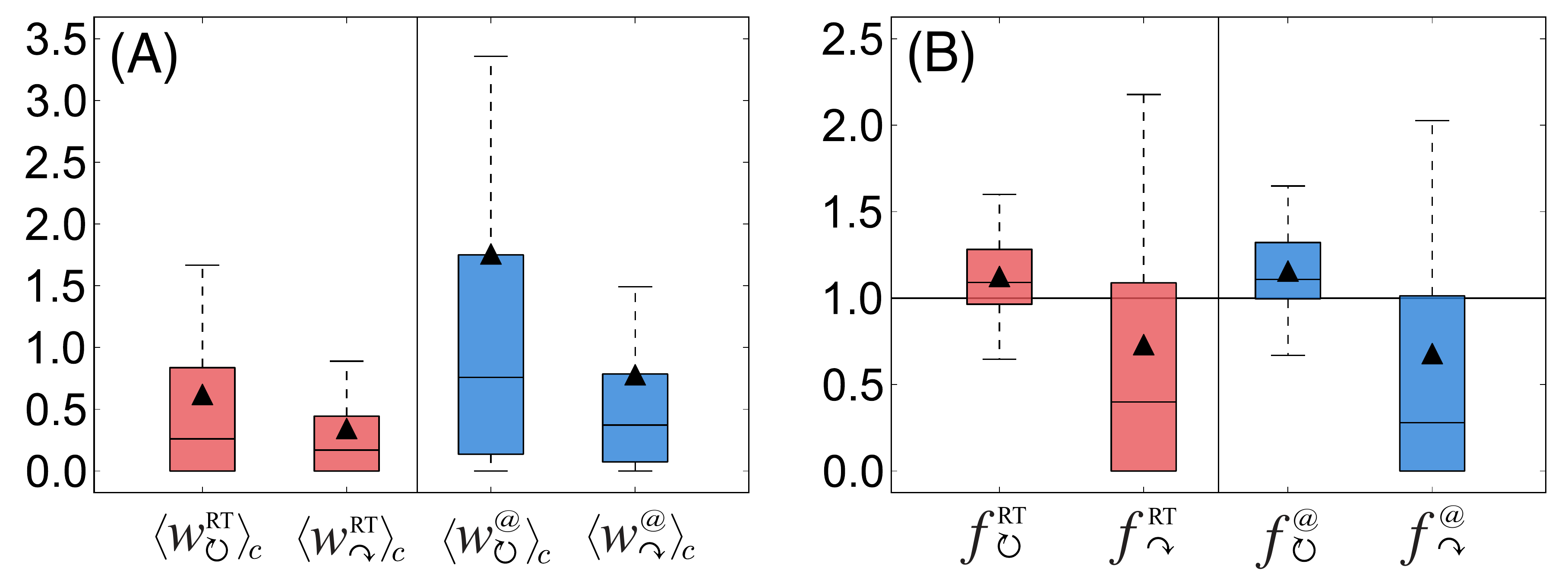}
\caption{Meme concentration in communities.  
We measure weights and focus in terms of retweets ($\mathrm{RT}$) or mentions ($\mathrm{@}$). 
We show (A)~\emph{community edge weight} and (B)~\emph{user community focus} 
using box plots.  Boxes cover 50\% of data and whisker cover 95\%.  The line and
triangle in a box represent the median and mean, respectively.
\label{fig:edge_weight}
} 
\end{figure}

Do memes spread like complex contagions in general? If social
reinforcement and homophily significantly influence the spread of memes, we
expect more communication within than across communities.  
Let us define the weight $w$ of an edge by the frequency of communication  
between the users connected by the edge.  
Nodes are partitioned into dense communities based on the structure of the 
network, but without knowledge of the weights (see Methods). 
For each community $c$, the average edge weights of intra- and
inter-community links, $\langle w_{\circlearrowright} \rangle_c$ and $\langle
w_{\curvearrowright} \rangle_c$, 
quantify how much information flows within and
across communities, respectively.  
We measure weights by aggregating all the meme spreading events in our data. If memes 
spread obliviously to community structure, like simple contagions, we would expect no
difference between intra- and inter-community links. 
By contrast, we observe that the intra-community links carry more
messages (Fig.~\ref{fig:edge_weight}(A)). Similar results have been reported from other datasets\cite{JP2007pnas,grabowicz2012social}.
%
In addition, by defining the focus of an individual as the fraction of activity that is directed to each neighbor in the same community, 
$f_{\circlearrowright}$, or in different communities, $f_{\curvearrowright}$,
we find that people interact more with members of the same community
(Fig.~\ref{fig:edge_weight}(B)). All the results are statistically significant ($p \ll 0.001$) and
robust across community detection methods (see Supplementary Information for additional details).

\subsection*{Meme Concentration in Communities} 

These results suggest that communities strongly trap communication. To quantify
this effect for individual memes, let us define the concentration of a meme in communities.
We expect more concentrated communication and meme adoption   
within communities if the meme spreads like a
complex contagion. To gauge this effect, we introduce four baseline models. 
The \emph{random sampling model} ($M_1$) assumes equal adoption probability for
everyone, ignoring network topology and all activity. 
The \emph{simple cascade model} ($M_2$) simulates the spreading of simple contagions\cite{karsai2011small}. 
The \emph{social reinforcement model} ($M_3$) employs a simple social reinforcement mechanism in addition to considering the network structure. 
In the \emph{homophily model} ($M_4$), users prefer to adopt the same ideas that are adopted by others in the same community. 
The simulation mechanisms of the four baseline models are summarized in Table~\ref{table:baselines}.

We estimate the trapping effects on memes by comparing the empirical data with these models. 
Note that we only focus on new memes (see definition in Methods). 
%
%
Let us define the concentration of a meme $h$ based on the
proportions of \emph{tweets} in each community. The usage-dominant community
$c^t(h)$ is the community generating most tweets with $h$.  The \emph{usage
dominance} of $h$, $r(h)$, is the proportion of tweets produced in the dominant
community $c^t(h)$ out of the total number of tweets $T(h)$ containing the
meme.  We also compute the \emph{usage entropy} $H^t(h)$ based on how tweets
containing $h$ are distributed across different communities. 
The relative usage dominance $r(h)/r_{M_1}(h)$ and entropy
$H^t(h)/H^t_{M_1}(h)$ are calculated using $M_1$ as baseline.
%
%
Analogous concentration measures can be defined based on \emph{users}. Let
$g(h)$ be the \emph{adoption dominance} of $h$, i.e., the proportion of the
$U(h)$ adopters in the community with most adopters. The \emph{adoption entropy}
$H^u(h)$ is computed based on how adopters of $h$ are allocated across
communities. 
The higher the dominance or the lower the entropy, the stronger the concentration 
of the meme. 
All measures are computed only based on tweets containing each meme in its early stage (first 50 tweets) 
to avoid any bias from the meme's popularity.

\begin{figure}[t]
\centering
\includegraphics[width=\columnwidth]{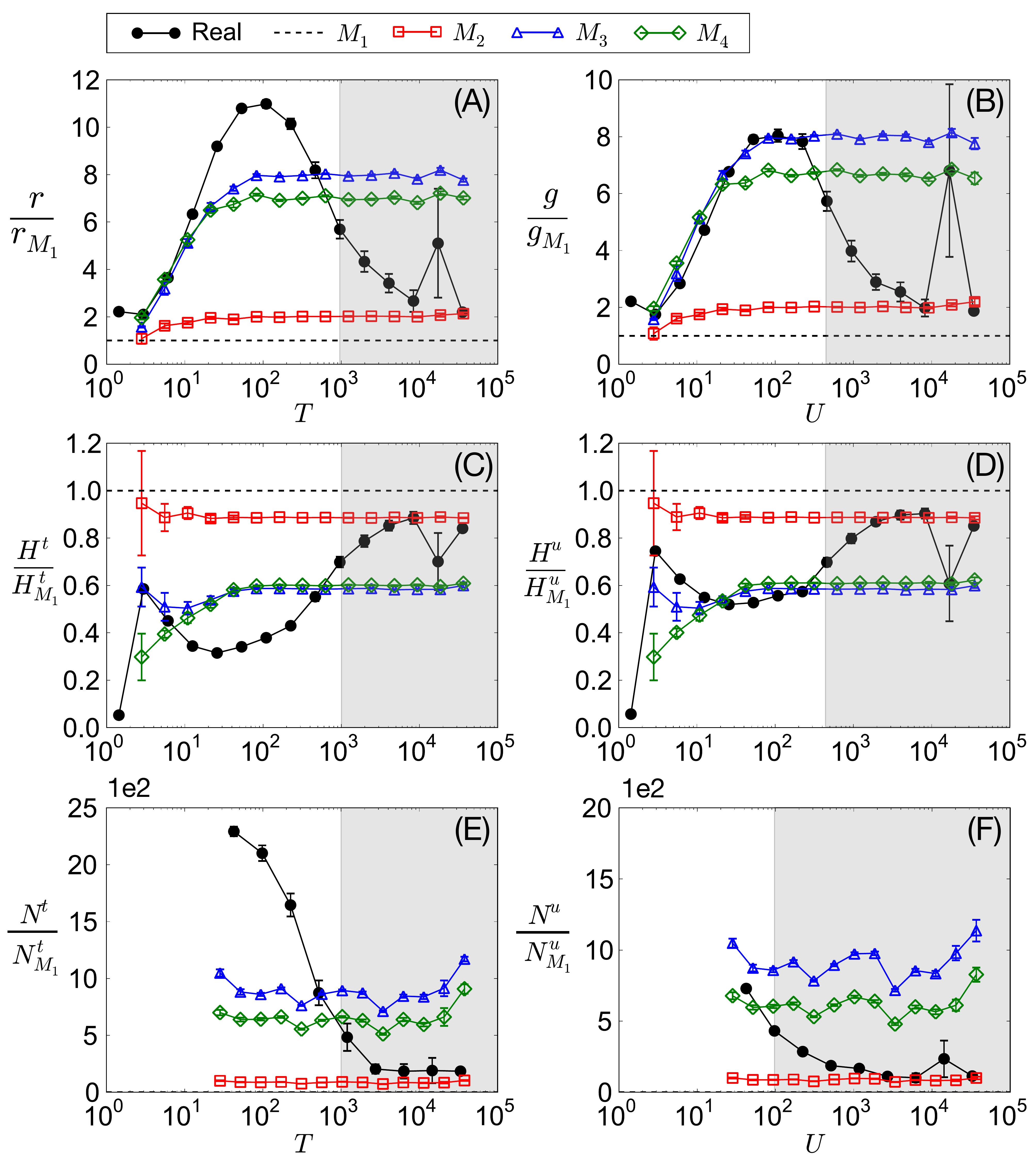}
\caption{Meme concentration in communities.
Changes in meme concentration as a function of meme popularity are illustrated by plotting relative 
(A)~\emph{usage dominance}, (B)~\emph{adoption dominance}, 
(C)~\emph{usage entropy}, and (D)~\emph{adoption entropy}.  
The relative dominance and entropy
ratios are averaged across hashtags in each popularity bin, with popularity
defined as number of tweets $T$ or adopters $U$; error bars indicate standard
errors.  Gray areas represent the ranges of popularity in which actual data
exhibit weaker concentration than both baseline models $M_3$ and $M_4$. 
The effect of multiple social reinforcement is estimated by \emph{average exposures} for every meme. The exposures can be measured in terms of (E) tweets or (F) users.
Similar results for different types of networks and community methods are described in SI.
\label{fig:concentration}
} 
\end{figure}

Figures~\ref{fig:concentration}(A-D) demonstrate that non-viral memes exhibit  concentration similar to (or stronger than) baselines $M_3$ or $M_4$, suggesting that these memes tend to spread like complex contagions. 
Note that models $M_2$, $M_3$, and $M_4$ produce stronger concentration than random sampling ($M_1$), because $M_2$ incorporates the structural trapping effect in simple cascades, $M_3$ considers both structural trapping and social reinforcement, and $M_4$ captures both structural trapping and homophily. 

Do \emph{all} memes spread like complex contagions? While the majority of memes are not viral, viral memes are adopted differently. Their concentration in the empirical data is the same as that of the simple cascade model $M_2$ (see the gray areas in Fig.~\ref{fig:concentration}(A-D)); community structure does not seem to trap successful memes as much as others.  These memes spread like simple contagions, permeating through many communities.

\subsection*{Strength of Social Reinforcement}

To further distinguish viral memes from others in terms of types of contagion, let us explicitly estimate the strength of social reinforcement. For a given meme $h$, we count the number of exposures that each adopter has experienced before the adoption and compute the \emph{average exposures} across all adopters, representing the strength of social reinforcement on $h$, labelled as $N(h)$. The exposures can be measured in terms of tweets $N^t(h)$ or users $N^u(h)$.
We compute relative average exposures, $N(h)/N_{M_1}(h)$, using only tweets at the early stages (first $50$ tweets). 
If this quantity is large, adoptions are more likely to happen with multiple social reinforcement and thus the meme spreads like a complex contagion. 
As shown in Fig.~\ref{fig:concentration}(E-F), viral memes require as little reinforcement as the simple cascade model $M_2$, while non-viral memes need as many exposures as $M_3$ or $M_4$. We arrive at the same conclusion: viral memes spread like simple contagions rather than like complex ones.

\subsection*{Prediction}

\begin{figure}
\centering
\includegraphics[width=\columnwidth]{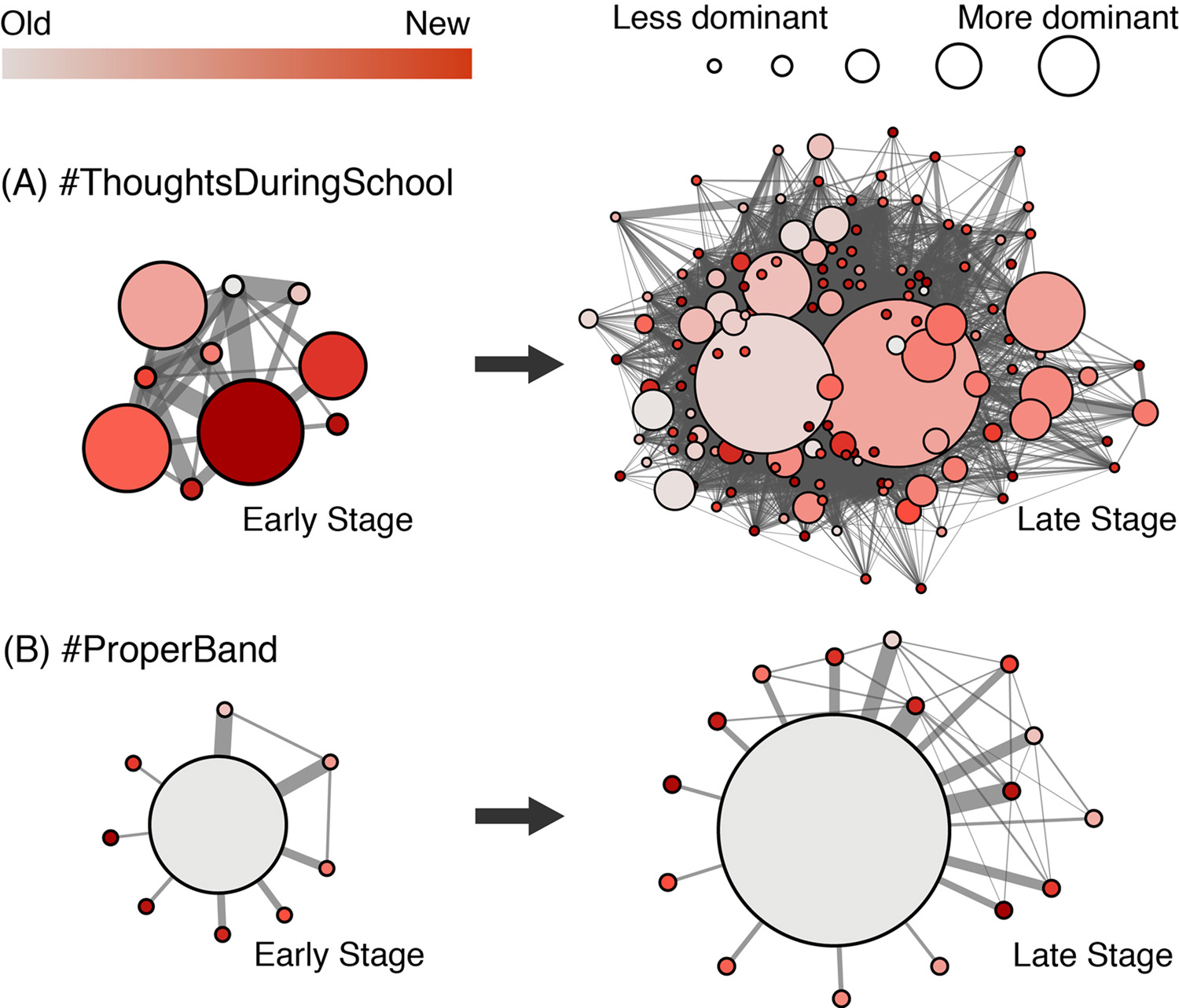}
\caption{
Evolution of two contrasting memes (viral vs. non-viral) in terms of community
structure. We represent each community as a node, whose size is proportional to
the number of tweets produced by the community. The color of a community
represents the time when the hashtag is first used in the community.
(A)~The evolution of a viral meme (\texttt{\#ThoughtsDuringSchool}) from the
early stage (30 tweets) to the late stage (200 tweets) of diffusion. (B)~The
evolution of a non-viral meme (\texttt{\#ProperBand}) from the early stage 
to the final stage (65 tweets).  
\label{fig:predict_viz}}
\end{figure}

The above findings imply an intriguing possibility: high concentration of a meme would hint that the meme is only interesting to certain communities, while weak concentration would imply a universal appeal and therefore might be used to predict the virality of the meme. 
To illustrate this intuition about the predictive power of the community structure, we show in
Fig.~\ref{fig:predict_viz} how the diffusion pattern of a viral meme differs from that of a
non-viral one, when analyzed through the lens of community concentration.

Let us therefore apply a machine learning technique, the \emph{random forests}
classification algorithm, 
to predict meme virality based on community concentration in the early diffusion stage. 
%
%
We employ two basic statistics based on early popularity and three types of community-based features in the prediction model, listed below. 

\begin{enumerate}

\item \textbf{Basic features based on early popularity.} Two basic statistical features are 
included in the prediction model. The number of \emph{early adopters} is the number 
of distinct users who generated the earliest tweets. The number of \emph{uninfected
neighbors of early adopters} characterizes the set of users who can adopt the
meme during the next step.

\item \textbf{Infected communities.} The simplest feature related to communities is 
the number of \emph{infected communities}, i.e., the number of communities containing early adopters. 

\item \textbf{Usage and adoption entropy.} $H^t(h)$ and $H^u(h)$ are good
indicators of the strength of meme concentration, as shown in Fig.~\ref{fig:concentration}.

\item \textbf{Fraction of intra-community user interactions}.  
We count pair-wise user interactions about any given meme, and calculate the proportion that occur between people in the same community.

\end{enumerate}

\begin{figure}
\centering
\includegraphics[width=\columnwidth]{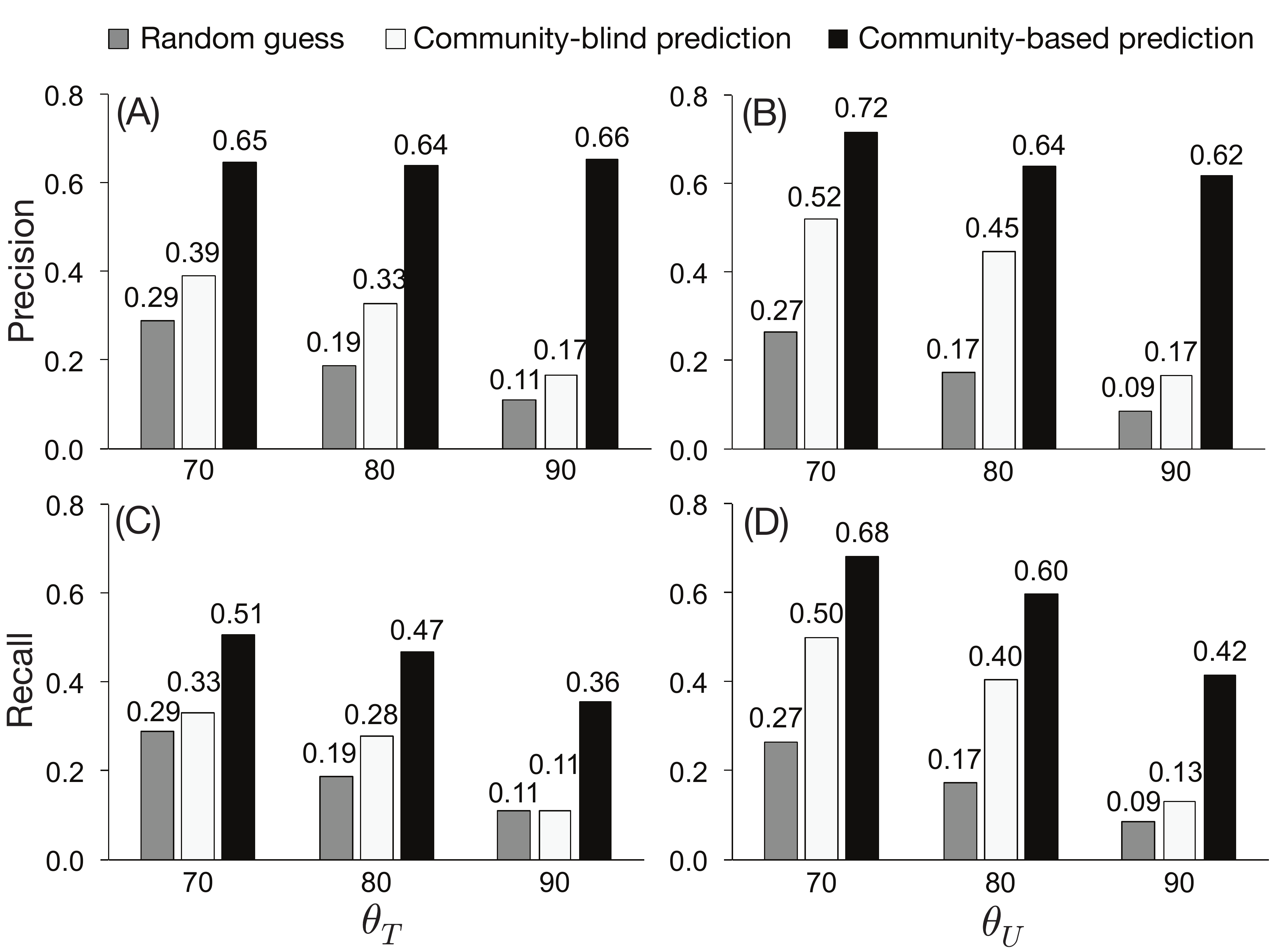}

\caption{Prediction performance. We predict whether a meme will go viral or
    not; a meme is labeled as \emph{viral} if it produces more tweets ($T$) or is adopted by more users ($U$) than a certain
    percentile threshold ($\theta = 70, 80, 90$) of memes.
    We use the random forests classifier trained on community concentration
    features, which are calculated based on the initial $n=50$ tweets for each
    meme. Prediction results are robust across different networks and community
    detection methods (see SI).  We compute precision
    and recall to compare our prediction results against two baselines.
 \label{fig:predict}}
\end{figure}

Our method aims to discover viral memes. To label viral memes, 
we rank all memes in our dataset based on numbers of tweets or adopters, and define a percentile threshold. A threshold of $\theta_T$ or $\theta_U$ means that a meme is deemed viral if it is mentioned in more tweets than $\theta_T$\% of the memes, or adopted by more users than $\theta_U$\% of the memes, respectively. 
All the features are computed based on the first
$50$ tweets for each hashtag $h$.  Two baselines are set up for comparison. 
\emph{Random guess} selects $n_{\mathrm{viral}}$  memes at random, where
$n_{\mathrm{viral}}$ is the number of viral memes in the actual data. 
\emph{Community-blind prediction} employs the same learning algorithm as ours but
without the community-based features. We compute both precision and recall
for evaluation; 
the former measures the proportion of predicted viral memes that are actually viral in the real data, 
and the latter quantifies how many of the viral memes are correctly predicted.  Our
community-based prediction excels in both precision and recall, indicating that
communities are helpful in capturing viral memes (Fig.~\ref{fig:predict}).  
For example, when detecting the most viral memes by users ($\theta_U = 90$), our method is about seven times as precise as random guess and over three times as precise as prediction without community features. We achieve a recall over 350\% better than random guess and over 200\% better than community-blind prediction. Similar results are obtained using different community detection methods or different types of social network links (see SI).

\section*{Discussion}

Despite the vast and growing literature on network communities, the importance
of community structure has not been fully explored and understood. 
%
%
Our findings expose an important role of community structure in the spreading of memes.
While the role of weak ties between different communities in
information diffusion has been recognized for decades\cite{weak_ties_1973,
JP2007pnas}, we provide a direct approach for translating data about 
community structure into predictive knowledge about what information will 
spread virally. Our method does not exploit message content, and can be 
easily applied to any socio-technical network from a small sample of data. 
This result can 
be relevant for 
online marketing and other social media applications.

%
Further analyses of network community structure in relation to social processes
hold potential for characterizing and forecasting social behavior. We
believe that many other complex dynamics of human society, from ethnic tension
to global conflicts, and from grassroots social movements to political
campaigns\cite{borge2011structural,conover2013ows,conover2013geospatial}, 
could be better understood by continued investigation of network structure.
%

\section*{Methods}

We collected a 10\% sample of all public tweets from Mar 24 to Apr 25, 2012 using the Twitter streaming API (\texttt{dev.twitter.com/docs/streaming-apis}). Only tweets written in English are extracted. The dataset comprises 121,807,378 tweets generated by 14,599,240 unique users, and containing at least one of 10,393,465 hashtags. We then constructed an undirected, unweighted network based on reciprocal following relationships between 595,460 randomly selected users, as bi-directional links reflect more stable and reliable social connections. Such a conservative choice to exclude information about direction and weights of links makes the approach more generally applicable to cases where static data about the social network is more readily available than dynamic data about information flow. Two other types of networks constructed on the basis of retweets and mentions were also tested for robustness (see extended analyses in SI).

We apply Infomap\cite{Rosvall2008InfoMap}, an established algorithm to identify
the community structure. To ensure the robustness of our results, we perform
the same analyses using another widely-used but very different community
detection method, link clustering\cite{Ahn2010LinkCommunity}.  The results are
similar (see details in SI). The network remains unweighted for community identification, 
to focus purely on the connection structure.

For quantifying meme concentration in communities and the strength of social reinforcement, we focus on \emph{new} memes that emerged during our observation time window. New memes are defined as those with fewer than 20 tweets during the previous month (Feb 24 -- Mar 23, 2012). A sensitivity test of our results with respect to hashtag filtering criteria is available in SI.

To replicate the Twitter API sampling effect in the baseline models, each simulation runs until 10 times more tweets are generated than the empirical numbers. Then, we select 10\% of the tweets at random. 
Every simulation is repeated 100 times and the 10\%-sampling is repeated 10 times on each simulation outcome. Thus, the average values of the measures from our toy models are computed across $100 \times 10$ samples.

In prediction, we use the random forest algorithm, an ensemble classifier that constructs 500 decision trees~\cite{Breiman:2001fk}. Each decision tree is trained with 4 random features independently and the final prediction outcomes combine the outputs of all the trees. The good performance of the random forest model benefits from the assumption that an ensemble of ``weak learners'' can form a ``strong learner.''
For training and testing, we employ 10-fold cross validation.



\begin{thebibliography}{10}
\expandafter\ifx\csname url\endcsname\relax
  \def\url#1{\texttt{#1}}\fi
\expandafter\ifx\csname urlprefix\endcsname\relax\def\urlprefix{URL }\fi
\providecommand{\bibinfo}[2]{#2}
\providecommand{\eprint}[2][]{\url{#2}}

\bibitem{newman_book}
\bibinfo{author}{Newman, M. E.~J.}
\newblock \emph{\bibinfo{title}{Networks: An Introduction}}
  (\bibinfo{publisher}{Oxford University Press}, \bibinfo{year}{2010}).

\bibitem{kleinberg_book}
\bibinfo{author}{Easley, D.} \& \bibinfo{author}{Kleinberg, J.}
\newblock \emph{\bibinfo{title}{Networks, Crowds, and Markets: Reasoning About
  a Highly Connected World}} (\bibinfo{publisher}{Cambridge University Press},
  \bibinfo{year}{2010}).

\bibitem{goffman1964nature}
\bibinfo{author}{Goffman, W.} \& \bibinfo{author}{Newill, V.~A.}
\newblock \bibinfo{title}{Generalization of epidemic theory: An application to
  the transmission of ideas}.
\newblock \emph{\bibinfo{journal}{Nature}} \textbf{\bibinfo{volume}{204}},
  \bibinfo{pages}{225---228} (\bibinfo{year}{1964}).

\bibitem{Daley1964rumour}
\bibinfo{author}{Daley, D.~J.} \& \bibinfo{author}{Kendall, D.~G.}
\newblock \bibinfo{title}{Epidemics and rumours}.
\newblock \emph{\bibinfo{journal}{Nature}} \textbf{\bibinfo{volume}{204}},
  \bibinfo{pages}{1118--1118} (\bibinfo{year}{1964}).

\bibitem{Anderson1992book}
\bibinfo{author}{Anderson, R.~M.}, \bibinfo{author}{May, R.~M.} \&
  \bibinfo{author}{Anderson, B.}
\newblock \emph{\bibinfo{title}{Infectious Diseases of Humans: Dynamics and
  Control}} (\bibinfo{publisher}{Oxford Science Publications},
  \bibinfo{year}{1992}).

\bibitem{Rogers:2003vn}
\bibinfo{author}{Rogers, E.}
\newblock \emph{\bibinfo{title}{Diffusion of Innovations}}
  (\bibinfo{publisher}{Free Press}, \bibinfo{year}{2003}),
  \bibinfo{edition}{5th} edn.

\bibitem{vesp2008book}
\bibinfo{author}{Barrat, A.}, \bibinfo{author}{Barth\'{e}lemy, M.} \&
  \bibinfo{author}{Vespignani, A.}
\newblock \emph{\bibinfo{title}{Dynamical Processes on Complex Networks}}
  (\bibinfo{publisher}{Cambridge University Press}, \bibinfo{year}{2008}).

\bibitem{vespignani2001SISonSF}
\bibinfo{author}{Pastor-Satorras, R.} \& \bibinfo{author}{Vespignani, A.}
\newblock \bibinfo{title}{Epidemic spreading in scale-free networks}.
\newblock \emph{\bibinfo{journal}{Phys. Rev. Lett.}}
  \textbf{\bibinfo{volume}{86}}, \bibinfo{pages}{3200--3203}
  (\bibinfo{year}{2001}).

\bibitem{Centola2010SpreadExp}
\bibinfo{author}{Centola, D.}
\newblock \bibinfo{title}{The spread of behavior in an online social network
  experiment}.
\newblock \emph{\bibinfo{journal}{Science}} \textbf{\bibinfo{volume}{329}},
  \bibinfo{pages}{1194--1197} (\bibinfo{year}{2010}).

\bibitem{ugander2012diversity}
\bibinfo{author}{Ugander, J.}, \bibinfo{author}{Backstrom, L.},
  \bibinfo{author}{Marlow, C.} \& \bibinfo{author}{Kleinberg, J.}
\newblock \bibinfo{title}{Structural diversity in social contagion}.
\newblock \emph{\bibinfo{journal}{Proc. Natl. Acad. Sci. USA}}
  \textbf{\bibinfo{volume}{109}}, \bibinfo{pages}{5962--5966}
  (\bibinfo{year}{2012}).

\bibitem{bond-facebook-2012}
\bibinfo{author}{Bond, R.~M.} \emph{et~al.}
\newblock \bibinfo{title}{A 61-million-person experiment in social influence
  and political mobilization}.
\newblock \emph{\bibinfo{journal}{Nature}} \textbf{\bibinfo{volume}{489}},
  \bibinfo{pages}{295--298} (\bibinfo{year}{2012}).

\bibitem{christ2007obesity}
\bibinfo{author}{Christakis, N.~A.} \& \bibinfo{author}{Fowler, J.~H.}
\newblock \bibinfo{title}{The spread of obesity in a large social network over
  32 years}.
\newblock \emph{\bibinfo{journal}{New England Journal of Medicine}}
  \textbf{\bibinfo{volume}{357}}, \bibinfo{pages}{370 -- 379}
  (\bibinfo{year}{2007}).

\bibitem{Vespignani2009}
\bibinfo{author}{Vespignani, A.}
\newblock \bibinfo{title}{Predicting the behavior of techno-social systems}.
\newblock \emph{\bibinfo{journal}{Science}} \textbf{\bibinfo{volume}{325}},
  \bibinfo{pages}{425--428} (\bibinfo{year}{2009}).

\bibitem{Lazer2009}
\bibinfo{author}{Lazer, D.} \emph{et~al.}
\newblock \bibinfo{title}{Computational social science}.
\newblock \emph{\bibinfo{journal}{Science}} \textbf{\bibinfo{volume}{323}},
  \bibinfo{pages}{721--723} (\bibinfo{year}{2009}).

\bibitem{moreno2004dynamics}
\bibinfo{author}{Moreno, Y.}, \bibinfo{author}{Nekovee, M.} \&
  \bibinfo{author}{Pacheco, A.~F.}
\newblock \bibinfo{title}{Dynamics of rumor spreading in complex networks}.
\newblock \emph{\bibinfo{journal}{Physical Review E}}
  \textbf{\bibinfo{volume}{69}}, \bibinfo{pages}{066130}
  (\bibinfo{year}{2004}).

\bibitem{Leskovec2009Meme}
\bibinfo{author}{Leskovec, J.}, \bibinfo{author}{Backstrom, L.} \&
  \bibinfo{author}{Kleinberg, J.}
\newblock \bibinfo{title}{Meme-tracking and the dynamics of the news cycle}.
\newblock In \emph{\bibinfo{booktitle}{Proc. ACM SIGKDD Intl. Conf. on
  Knowledge discovery and data mining}}, \bibinfo{pages}{497--506}
  (\bibinfo{year}{2009}).

\bibitem{borge2011structural}
\bibinfo{author}{Borge-Holthoefer, J.} \emph{et~al.}
\newblock \bibinfo{title}{Structural and dynamical patterns on online social
  networks: the spanish may 15th movement as a case study}.
\newblock \emph{\bibinfo{journal}{PLoS One}} \textbf{\bibinfo{volume}{6}},
  \bibinfo{pages}{e23883} (\bibinfo{year}{2011}).

\bibitem{lehmann2012dynamical}
\bibinfo{author}{Lehmann, J.}, \bibinfo{author}{Gon{\c{c}}alves, B.},
  \bibinfo{author}{Ramasco, J.~J.} \& \bibinfo{author}{Cattuto, C.}
\newblock \bibinfo{title}{Dynamical classes of collective attention in
  twitter}.
\newblock In \emph{\bibinfo{booktitle}{Proceedings of the 21st international
  conference on World Wide Web}}, \bibinfo{pages}{251--260}
  (\bibinfo{organization}{ACM}, \bibinfo{year}{2012}).

\bibitem{aral2011viral}
\bibinfo{author}{Aral, S.} \& \bibinfo{author}{Walker, D.}
\newblock \bibinfo{title}{Creating social contagion through viral product
  design: A randomized trial of peer influence in networks}.
\newblock \emph{\bibinfo{journal}{Management Science}}
  \textbf{\bibinfo{volume}{57}}, \bibinfo{pages}{1623--1639}
  (\bibinfo{year}{2011}).

\bibitem{Leskovec2007viral}
\bibinfo{author}{Leskovec, J.}, \bibinfo{author}{Adamic, L.} \&
  \bibinfo{author}{Huberman, B.}
\newblock \bibinfo{title}{The dynamics of viral marketing}.
\newblock \emph{\bibinfo{journal}{ACM Trans. Web}} \textbf{\bibinfo{volume}{1}}
  (\bibinfo{year}{2007}).

\bibitem{Weng:2012scirep}
\bibinfo{author}{Weng, L.}, \bibinfo{author}{Flammini, A.},
  \bibinfo{author}{Vespignani, A.} \& \bibinfo{author}{Menczer, F.}
\newblock \bibinfo{title}{Competition among memes in a world with limited
  attention}.
\newblock \emph{\bibinfo{journal}{Scientific Reports}}
  \textbf{\bibinfo{volume}{2}(\bibinfo{number}{335})}, \bibinfo{doi}{10.1038/srep00335} (\bibinfo{year}{2012}).

\bibitem{Berger:2009viral}
\bibinfo{author}{Berger, J.} \& \bibinfo{author}{Milkman, K.~L.}
\newblock \bibinfo{title}{What makes online content viral?}
\newblock \emph{\bibinfo{journal}{Journal of Marketing Research}}
  \textbf{\bibinfo{volume}{49}}, \bibinfo{pages}{192--205}
  (\bibinfo{year}{2009}).

\bibitem{Jamali2009digging}
\bibinfo{author}{Jamali, S.} \& \bibinfo{author}{Rangwala, H.}
\newblock \bibinfo{title}{Digging digg: Comment mining, popularity prediction,
  and social network analysis}.
\newblock In \emph{\bibinfo{booktitle}{Proc. Intl. Conf. on Web Information
  Systems and Mining (WISM)}}, \bibinfo{pages}{32 -- 38}
  (\bibinfo{year}{2009}).

\bibitem{Szabo2010pred}
\bibinfo{author}{Szabo, G.} \& \bibinfo{author}{Huberman, B.~A.}
\newblock \bibinfo{title}{Predicting the popularity of online content}.
\newblock \emph{\bibinfo{journal}{Communications of the ACM}}
  \textbf{\bibinfo{volume}{53}}, \bibinfo{pages}{80--88}
  (\bibinfo{year}{2010}).

\bibitem{Suh2010}
\bibinfo{author}{Suh, B.}, \bibinfo{author}{Hong, L.},
  \bibinfo{author}{Pirolli, P.} \& \bibinfo{author}{Chi, E.~H.}
\newblock \bibinfo{title}{Want to be retweeted? large scale analytics on
  factors impacting retweet in twitter network}.
\newblock In \emph{\bibinfo{booktitle}{Proc. IEEE Intl. Conf. on Social
  Computing}}, \bibinfo{pages}{177--184} (\bibinfo{year}{2010}).

\bibitem{Watts2002}
\bibinfo{author}{Watts, D.~J.}
\newblock \bibinfo{title}{A simple model of global cascades on random
  networks}.
\newblock \emph{\bibinfo{journal}{Proc. of the National Academy of Sciences}}
  (\bibinfo{year}{2002}).

\bibitem{Dodds2004}
\bibinfo{author}{Dodds, P.~S.} \& \bibinfo{author}{Watts, D.~J.}
\newblock \bibinfo{title}{Universal behavior in a generalized model of
  contagion}.
\newblock \emph{\bibinfo{journal}{Phys. Rev. Lett.}}
  \textbf{\bibinfo{volume}{92}}, \bibinfo{pages}{218701}
  (\bibinfo{year}{2004}).

\bibitem{Kitsak2010kcore}
\bibinfo{author}{Kitsak, M.} \emph{et~al.}
\newblock \bibinfo{title}{Identification of influential spreaders in complex
  networks}.
\newblock \emph{\bibinfo{journal}{Nature Physics}}
  \textbf{\bibinfo{volume}{6}}, \bibinfo{pages}{888--893}
  (\bibinfo{year}{2010}).

\bibitem{Backstrom2006GFL}
\bibinfo{author}{Backstrom, L.}, \bibinfo{author}{Huttenlocher, D.},
  \bibinfo{author}{Kleinberg, J.} \& \bibinfo{author}{Lan, X.}
\newblock \bibinfo{title}{Group formation in large social networks: membership,
  growth, and evolution}.
\newblock In \emph{\bibinfo{booktitle}{Proc. ACM SIGKDD Intl. Conf. on
  Knowledge discovery and data mining}}, \bibinfo{pages}{44--54}
  (\bibinfo{year}{2006}).

\bibitem{fortunato2010community}
\bibinfo{author}{Fortunato, S.}
\newblock \bibinfo{title}{Community detection in graphs}.
\newblock \emph{\bibinfo{journal}{Physics Reports}}
  \textbf{\bibinfo{volume}{486}}, \bibinfo{pages}{75--174}
  (\bibinfo{year}{2010}).

\bibitem{Rosvall2008InfoMap}
\bibinfo{author}{Rosvall, M.} \& \bibinfo{author}{Bergstrom, C.~T.}
\newblock \bibinfo{title}{Maps of random walks on complex networks reveal
  community structure}.
\newblock \emph{\bibinfo{journal}{Proc. Natl. Acad. Sci. USA}}
  \textbf{\bibinfo{volume}{105}}, \bibinfo{pages}{1118--1123}
  (\bibinfo{year}{2008}).

\bibitem{Ahn2010LinkCommunity}
\bibinfo{author}{Ahn, Y.-Y.}, \bibinfo{author}{Bagrow, J.} \&
  \bibinfo{author}{Lehmann, S.}
\newblock \bibinfo{title}{Link communities reveal multiscale complexity in
  networks}.
\newblock \emph{\bibinfo{journal}{Nature}} \textbf{\bibinfo{volume}{466}},
  \bibinfo{pages}{761--764} (\bibinfo{year}{2010}).

\bibitem{galstyan2007cascading}
\bibinfo{author}{Galstyan, A.} \& \bibinfo{author}{Cohen, P.}
\newblock \bibinfo{title}{Cascading dynamics in modular networks}.
\newblock \emph{\bibinfo{journal}{Physical Review E}}
  \textbf{\bibinfo{volume}{75}}, \bibinfo{pages}{036109}
  (\bibinfo{year}{2007}).

\bibitem{gleeson2008cascades}
\bibinfo{author}{Gleeson, J.~P.}
\newblock \bibinfo{title}{Cascades on correlated and modular random networks}.
\newblock \emph{\bibinfo{journal}{Physical Review E}}
  \textbf{\bibinfo{volume}{77}}, \bibinfo{pages}{046117}
  (\bibinfo{year}{2008}).

\bibitem{JP2007pnas}
\bibinfo{author}{Onnela, J.-P.} \emph{et~al.}
\newblock \bibinfo{title}{Structure and tie strengths in mobile communication
  networks}.
\newblock \emph{\bibinfo{journal}{Proc. Natl. Acad. Sci. USA}}
  \textbf{\bibinfo{volume}{104}}, \bibinfo{pages}{7332--7336}
  (\bibinfo{year}{2007}).

\bibitem{weak_ties_1973}
\bibinfo{author}{Granovetter, M.~S.}
\newblock \bibinfo{title}{The strength of weak ties}.
\newblock \emph{\bibinfo{journal}{American Journal of Sociology}}
  \textbf{\bibinfo{volume}{78}}, \bibinfo{pages}{1360--1379}
  (\bibinfo{year}{1973}).

\bibitem{grabowicz2012social}
\bibinfo{author}{Grabowicz, P.~A.}, \bibinfo{author}{Ramasco, J.~J.},
  \bibinfo{author}{Moro, E.}, \bibinfo{author}{Pujol, J.~M.} \&
  \bibinfo{author}{Eguiluz, V.~M.}
\newblock \bibinfo{title}{Social features of online networks: The strength of
  intermediary ties in online social media}.
\newblock \emph{\bibinfo{journal}{PloS one}} \textbf{\bibinfo{volume}{7}},
  \bibinfo{pages}{e29358} (\bibinfo{year}{2012}).

\bibitem{threshold_model}
\bibinfo{author}{Granovetter, M.~S.}
\newblock \bibinfo{title}{Threshold models of collective behavior}.
\newblock \emph{\bibinfo{journal}{American Journal of Sociology}}
  \textbf{\bibinfo{volume}{83}}, \bibinfo{pages}{1420--1433}
  (\bibinfo{year}{1978}).

\bibitem{McPherson2001homophily}
\bibinfo{author}{McPherson, M.}, \bibinfo{author}{Lovin, L.} \&
  \bibinfo{author}{Cook, J.}
\newblock \bibinfo{title}{Birds of a feather: Homophily in social networks}.
\newblock \emph{\bibinfo{journal}{Annual Review of Sociology}}
  \textbf{\bibinfo{volume}{27}}, \bibinfo{pages}{415--444}
  (\bibinfo{year}{2001}).

\bibitem{Centola2011homophily}
\bibinfo{author}{Centola, D.}
\newblock \bibinfo{title}{An experimental study of homophily in the adoption of
  health behavior}.
\newblock \emph{\bibinfo{journal}{Science}} \textbf{\bibinfo{volume}{334}},
  \bibinfo{pages}{1269--1272} (\bibinfo{year}{2011}).

\bibitem{aral_distinguishing_2009}
\bibinfo{author}{Aral, S.}, \bibinfo{author}{Muchnik, L.} \&
  \bibinfo{author}{Sundararajan, A.}
\newblock \bibinfo{title}{Distinguishing influence-based contagion from
  homophily-driven diffusion in dynamic networks}.
\newblock \emph{\bibinfo{journal}{Proc. Natl. Acad. Sci. USA}}
  \textbf{\bibinfo{volume}{106}}, \bibinfo{pages}{21544--21549}
  (\bibinfo{year}{2009}).

\bibitem{Shalizi2012homophily}
\bibinfo{author}{Shalizi, C.} \& \bibinfo{author}{Thomas, A.}
\newblock \bibinfo{title}{Homophily and contagion are generically confounded in
  observational social network studies}.
\newblock \emph{\bibinfo{journal}{Sociological Methods \& Research}}
  \textbf{\bibinfo{volume}{40}}, \bibinfo{pages}{211--239}
  (\bibinfo{year}{2011}).

\bibitem{karsai2011small}
\bibinfo{author}{Karsai, M.} \emph{et~al.}
\newblock \bibinfo{title}{Small but slow world: How network topology and
  burstiness slow down spreading}.
\newblock \emph{\bibinfo{journal}{Physical Review E}}
  \textbf{\bibinfo{volume}{83}}, \bibinfo{pages}{025102}
  (\bibinfo{year}{2011}).

\bibitem{conover2013ows}
\bibinfo{author}{Conover, M.~D.}, \bibinfo{author}{Ferrara, E.},
  \bibinfo{author}{Menczer, F.} \& \bibinfo{author}{Flammini, A.}
\newblock \bibinfo{title}{The digital evolution of occupy wall street}.
\newblock \emph{\bibinfo{journal}{{PLOS ONE}}} \textbf{\bibinfo{volume}{8}},
  \bibinfo{pages}{e64679} (\bibinfo{year}{2013}).

\bibitem{conover2013geospatial}
\bibinfo{author}{Conover, M.~D.} \emph{et~al.}
\newblock \bibinfo{title}{The geospatial characteristics of a social movement
  communication network}.
\newblock \emph{\bibinfo{journal}{PloS one}} \textbf{\bibinfo{volume}{8}},
  \bibinfo{pages}{e55957} (\bibinfo{year}{2013}).

\bibitem{Breiman:2001fk}
\bibinfo{author}{Breiman, L.}
\newblock \bibinfo{title}{Random forests}.
\newblock \emph{\bibinfo{journal}{Machine Learning}}
  \textbf{\bibinfo{volume}{45}}, \bibinfo{pages}{5--32} (\bibinfo{year}{2001}).

\bibitem{romero2013topic}
\bibinfo{author}{Romero, D.~M.}, \bibinfo{author}{Tan, C.} \&
  \bibinfo{author}{Ugander, J.}
\newblock \bibinfo{title}{On the interplay between social and topical
  structure}.
\newblock In \emph{\bibinfo{booktitle}{Proc. AAAI Intl. Conf. on Weblogs and
  Social Media}} (\bibinfo{year}{2013}).

\end{thebibliography}

\newpage

\section*{Acknowledgments}

We thank Albert-L\'{a}szl\'{o} Barab\'{a}si, Haewook Kwak, James P. Bagrow, 
Cosma Rohilla Shalizi, Yang-Yu Liu, Sune Lehmann, and Sue Moon for helpful 
discussions and comments on earlier versions of this manuscript, and Yangyi 
Chen for help on using FutureGrid clusters for processing large-scale data.
This manuscript is based upon work supported in part by the James S. 
McDonnell Foundation and the National Science Foundation under Grants 
No. 0910812 and 1101743. 
While finalizing this manuscript, we have been made aware of a related paper developed independently by D.R. Romero, C. Tan, and J. Ugander~\cite{romero2013topic}.

\section*{Author Contributions}

L.W. and Y.-Y.A designed the study. L.W. performed data collection and measurements. L.W., F.M., and Y.-Y.A contributed to analysis and manuscript preparation.

\section*{Additional Information}

\begin{description}
\item[Competing Interests] Y.-Y.A. declares competing financial interests: he co-founded a social media analytics company ``Janys Analytics'' in 2011 and is currently one of the major shareholders. The other authors declare no competing financial interests.  
\item[Correspondence] Correspondence and requests for materials should be addressed to \url{yyahn@indiana.edu}.
\end{description}

\newpage

\begin{table}
\caption{Baseline models for information diffusion.}
\centering
\begin{tabular}{ l | c c c | p{25em} }
\hline
 & \multicolumn{3}{c|}{Community effects} & \multirow{2}{*}{Simulation implementation} \\
 & {\small Network} & {\small Reinforcement} & {\small Homophily} & \\
 \hline\hline
$M_1$ &  &  &  & For a given hashtag $h$, $M_1$ randomly samples the same number of tweets or users as in the real data. \\
$M_2$ & $\checkmark$ & & & $M_2$ takes the network structure into account while neglecting social reinforcement and homophily. $M_2$ starts with a random seed user. At each step, with probability $p$, an infected node is randomly selected and one of its neighbors adopts the meme, or with probability $1-p$, the process restarts from a new seed user ($p = 0.85$). \\
$M_3$ & $\checkmark$ & $\checkmark$ & & The cascade in $M_3$ is generated similarly to $M_2$ but at each step the user with the maximum number of infected neighbors adopts the meme. 
\\
$M_4$ & $\checkmark$ & & $\checkmark$ & In $M_4$, the simple cascading process is simulated in the same way as in $M_2$ but subject to the constraint that at each step, only neighbors in the same community have a chance to adopt the meme.  
\\
\hline
\end{tabular}
\label{table:baselines}
\end{table}%

\end{document}